\documentclass[sigconf]{acmart}

\AtBeginDocument{%
  }

\copyrightyear{2024}
\acmYear{2024}
\setcopyright{rightsretained}
\acmConference[FORGE '24]{AI Foundation Models and Software
Engineering}{April 14, 2024}{Lisbon, Portugal}
\acmBooktitle{AI Foundation Models and Software Engineering (FORGE '24), April
14, 2024, Lisbon, Portugal}\acmDOI{10.1145/3650105.3652302}
\acmISBN{979-8-4007-0609-7/24/04}

\usepackage{listings}
\usepackage{multirow}
\usepackage{booktabs}
\usepackage{makecell}

\acmISBN{978-1-4503-XXXX-X/18/06}

\begin{document}
\sloppy

\title{Creative and Correct:\\Requesting Diverse Code Solutions from AI Foundation Models}

\author{Scott Blyth}
\email{sbly0002@student.moansh.edu}
\affiliation{
  \institution{Monash University}
  \city{Clayton}
  \country{Australia}
}
\author{Markus Wagner}
\email{markus.wagner@monash.edu}
\affiliation{
  \institution{Monash University}
  \city{Clayton}
  \country{Australia}
}
\author{Christoph Treude}
\email{ctreude@smu.edu.sg}
\affiliation{
  \institution{Singapore Management University}
  \city{Singapore}
  \country{Singapore}
}
\begin{abstract}

AI foundation models have the capability to produce a wide array of responses to a single prompt, a feature that is highly beneficial in software engineering to generate diverse code solutions. However, this advantage introduces a significant trade-off between diversity and correctness. In software engineering tasks, diversity is key to exploring design spaces and fostering creativity, but the practical value of these solutions is heavily dependent on their correctness. Our study systematically investigates this trade-off using experiments with HumanEval tasks, exploring various parameter settings and prompting strategies. We assess the diversity of code solutions using similarity metrics from the code clone community. The study identifies combinations of parameters and strategies that strike an optimal balance between diversity and correctness, situated on the Pareto front of this trade-off space. These findings offer valuable insights for software engineers on how to effectively use AI foundation models to generate code solutions that are diverse and accurate.


\end{abstract}



\maketitle


\section{Introduction}

The role of diversity in programming solutions is increasingly recognised as essential for innovation and problem solving in software engineering~\cite{baudry2015multiple}. The need for diversity in programming solutions is rooted in two main reasons. First, the selection of the right solution for a programming task depends on the context, which involves factors such as project requirements, the skills of the development team, the environment for deployment, and key aspects such as security, performance, and maintainability~\cite{traini2021software}. For example, in high-performance scenarios, such as real-time systems, the focus might shift from code that is easy to read and maintain to one that delivers on performance~\cite{treude2023navigating}.

Second, being aware of diverse solutions is important in encouraging creativity and new ideas in software engineering~\cite{treude2022taming}. Presenting different solutions helps to address ambiguous challenges and promotes creative thinking and serendipity, while also avoiding repetitive or limited approaches~\cite{ziarani2021serendipity}. 

However, it is important to maintain a balance between this diversity and the correctness of the solutions, as too much diversity might lead to impractical or incorrect outcomes. The specific challenge addressed in this paper is how best to utilise AI foundation models to generate a wide array of programming solutions that not only exhibit diversity but are also correct. Our exploration includes adjusting parameters and modifying the prompts in these models to broaden the spectrum of viable solutions while ensuring that quality is not compromised.

Our approach involves systematic experimentation with HumanEval tasks, using similarity metrics from the code clone community~\cite{sajnani2016sourcerercc} to evaluate the diversity of code solutions generated. We find that employing multiple rounds of prompts is an effective method for generating code solutions that are both diverse and accurate. In addition, integrating various strategies, such as adjusting frequency penalties, moves the Pareto front toward more optimal trade-offs between diversity and accuracy, demonstrating the potential of nuanced model configuration to achieve superior results in code generation. These findings aim to help software engineers effectively use AI foundation models to create code solutions that are diverse, accurate, and practical. Our scripts and data are available at \url{https://github.com/scottb341/diverse\_code\_generator}.













\section{HumanEval Tasks}\label{sec:humaneval}

We experiment with the HumanEval task dataset \cite{HumanEval}, which is a set of 164 tasks used to test the performance of foundation models on programming tasks. 

Each task has a description and a set of test cases (average of 8.08 per task). The description of a task contains the function signature (name, type hinting) and some example test cases. As an example, here is the description for Task~4 in the HumanEval dataset:

\begin{lstlisting}[
    basicstyle=\footnotesize, %or \tiny or \small or \footnotesize etc.
]
from typing import List

def mean_absolute_deviation(numbers: List[float]) -> float:
    """ For a given list of input numbers, calculate Mean 
    Absolute Deviation around the mean of this dataset.
    Mean Absolute Deviation is the average absolute 
    difference between each element and a centerpoint 
    (mean in this case):
    MAD = average | x - x_mean |
    >>> mean_absolute_deviation([1.0, 2.0, 3.0, 4.0])
    1.0
    """
\end{lstlisting}

For this study, we use different strategies and parameter settings to create ten code solutions for each HumanEval task, with the goal of exploring the trade-off between diversity and accuracy. 

We call the response of a foundation model to a task a ``(solution) candidate''. We do this to distinguish it from an actual solution that could be considered functionally correct with respect to the specified task.


\section{Diversity Assessment}

To assess how diverse the solutions are, we employ clone detectors. 
Code clones refer to code snippets that have similar functionality~\cite{dangClone}. 
In our study, we assess diversity using the SourcererCC tool~\cite{sajnani2016sourcerercc} and using the cosine similarity of the vector embeddings.

SourcererCC is a token-based, Type~3 \cite{dangClone} code clone detector: it can detect
``near miss clones'', which are syntactically similar except for the addition or removal of statements and the changes in comments, identifiers, types, literals, and layouts; therefore, they tend to be functionally similar. 
When presented with a set of candidates to analyse, SourcererCC reports the IDs of the pairs of clones. The granularity of this clone detector is functional, that is, method by method. To measure the similarity score (which we call sccSim) for a set of solutions for a task, we calculate the ratio of the number of detected clones to the number of possible clones; this means that if all code solutions are identical clones, then sccSim=1.0, and if all code solutions are pairwise different, then sccSim=0.0.

As an alternative to a token-based tool, we measure the cosine similarity of candidates in the vector embedding of a model trained on the CodeSearchNet~\cite{husain_codesearchnet_2019} dataset, CodeBERTaPy~\cite{CodeBertaPy}. This is similar to the approaches taken by Chung et al.~\cite{chung2023increasing} and Cevoli et al.~\cite{cevoli2021semantic}, who used vector embeddings to measure the linguistic similarity/diversity of generated texts. 
In our case, to calculate the similarity of two candidates $x$ and $y$, using the embedding from Code~BERTaPy, we compute the cosine of the vectors $\Vec{u}$ and $\Vec{v}$, where $\Vec{u}$ and $\Vec{v}$ are the vector embeddings for $x$ and $y$, respectively. 
When given a set of two or more candidates, we calculate all pairwise similarities and then return as the embedding-based similarity (which we call cosineSim) the mean over all pairwise results. 


\section{Correctness Assessment}

To determine the correctness of the code generated by foundation models, we employ the \textit{pass@k} metric as defined by Chen et al.~\cite{chen2021codex}. They measure performance by generating \textit{k} solution candidate samples, and if one candidate passes all test cases, then the foundation model is considered to have ``passed'' the task for that value of $k$; this means that, (1) in the case of $k=1$, the model gets one shot at passing the task and (2) in the case of $k>1$, a single functionally correct candidate (out of multiple) suffices to pass the task. 

However, computing pass@k in this way can have a high variance, depending on which subset of $k$ samples from a total of $n$ returned code solutions is considered. Instead, just like Chen et al.~\cite{chen2021codex}, to evaluate pass@k, we generate $n \geq k$ candidate solutions, count the number of correct samples $c \leq n$ which pass the unit tests, and then calculate the unbiased estimator 
(an estimate for the ratio passed problem to total problems):
$$ pass@k = \mathbb{E}_{Task} \left [ 1 - \frac{\begin{pmatrix} n-c \\ k \end{pmatrix}}{\begin{pmatrix} n \\ k \end{pmatrix}}  \right ]$$

In our analyses, we focus almost exclusively on pass@1 scores for two reasons: (1) no approach consistently gives us the required 10 solutions, thus preventing reliable comparisons at the pass@10 level and also below; (2) it is difficult to interpret pass@k scores in our context when $k$ is neither 1 (which shows how many correct code solutions are produced) nor 10 (which is our target number). 



\section{Configuring Foundation Models}

While a plethora of foundation models are available, we consider (due to practical reasons) only gpt-3.5 
and gpt-4 from OpenAI~\cite{openaiDocumentation}. 

The operation of both foundation models can be influenced by a number of configuration parameters. 
In the following, we describe the parameters, list their default values, and outline their intended effects.  


\paragraph{temperature}\label{sec:temperarture}
Temperature is used to control the randomness of the output of the foundation model. It is a value from 0 to 2, where 0 is the most deterministic and 2 is the most random. The default value is 1.

\paragraph{top\_p}\label{scc:top_p}
The top\_p parameter is a value from 0 (exclusive) to 1 (inclusive, default) that controls what the next generated token is pooled from. A top\_p value of 0.5 means that only the top 50\% of the tokens is considered and the bottom 50\% is discarded.

\paragraph{frequency\_penalty}\label{sec:frequency_pen}

Changes the probability that a token appears on the basis of its frequency. It ranges from -2 to 2, where a positive value reduces (i.e., penalises) the probability and a negative value increases the probability. The default value is 0.

\paragraph{presence\_penalty}\label{sec:pres_pen}
Similar to frequency\_penalty, the presence penalty modifies the probability of each token based on its prior appearance in the text. This means that a token already used is ``penalised'' or ``encouraged'' based on its individual occurrence. The values for this penalty range from -2 to 2, with positive values decreasing the probability of a token reappearing and negative values increasing it. The default value is 0, which implies that there is no adjustment for token repetition.

\paragraph{logit bias} 
The logit bias parameter is used to decrease the probability that the foundation model generates certain tokens. 
Chung et al.~\cite{chung2023increasing} found that the logit bias parameter was successful in increasing the linguistic diversity of the responses. 
Our process of mapping solutions to a logit bias dictionary follows Chung et al.: first, the top 100 tokens are taken from the previously generated solutions (the concatenation of all of the solutions), and then, based on their counts and the total number of tokens, the bias is computed for those tokens. Specifically, the bias for the token ``TOKEN'' is:

$$ Bias [TOKEN] = - max\_bias \cdot \frac{COUNT[TOKEN]}{num\_tokens} $$

where num\_tokens is the total number of tokens in the text that are in the top 100. We employ a maximum bias of 7.5 as seen in Chung et al.

\section{Prompting Techniques}

As an alternative to changing the operations of a foundation model by changing its operational configuration, we investigate different approaches to prompting it. Depending on the method, this can give the model extra context for its statistical inference, allowing it to find different solutions.

\paragraph{Regeneration}\label{sec:regen_prompt}

Our regeneration approach (\textit{regen}) is a simple process of giving the description of the task to the model and then repeating this $n$ times to produce $n$ solution candidates. Note that it suffices to hand over the HumanEval tasks (see Section~\ref{sec:humaneval} for an example), since they provide Python code as context request to the task. 
The prompts for generating two solutions for a task would be:

\small{
\vspace{2mm}
\noindent\texttt{--- establish a new connection to the model ---}\\
\texttt{Prompt: <task description>}\\
\texttt{Response: <candidate\_1>}\\
\texttt{--- establish a new connection to the model ---}\\
\texttt{Prompt: <task description>}\\
\texttt{Response: <candidate\_2>}\\
}

\paragraph{n\_different}

The n\_different prompt asks the foundation model for $n$ different solutions:

\texttt{<task description> + ``\textbackslash n Give me n different solutions for this problem in Python.}

Due to the maximum number of tokens that can be generated, this prompt has an upper bound on the number of candidates that can be returned. The maximum number of tokens is specified by the parameter \textit{max\_tokens} which is capped by the model's context length, which for gpt-3.5 is 4096 tokens. A value of $n=10$ is used for this paper.

\paragraph{n\_k\_different}

\begin{figure}
    \centering
    \includegraphics[width=8cm,trim={10 20 10 10},clip]{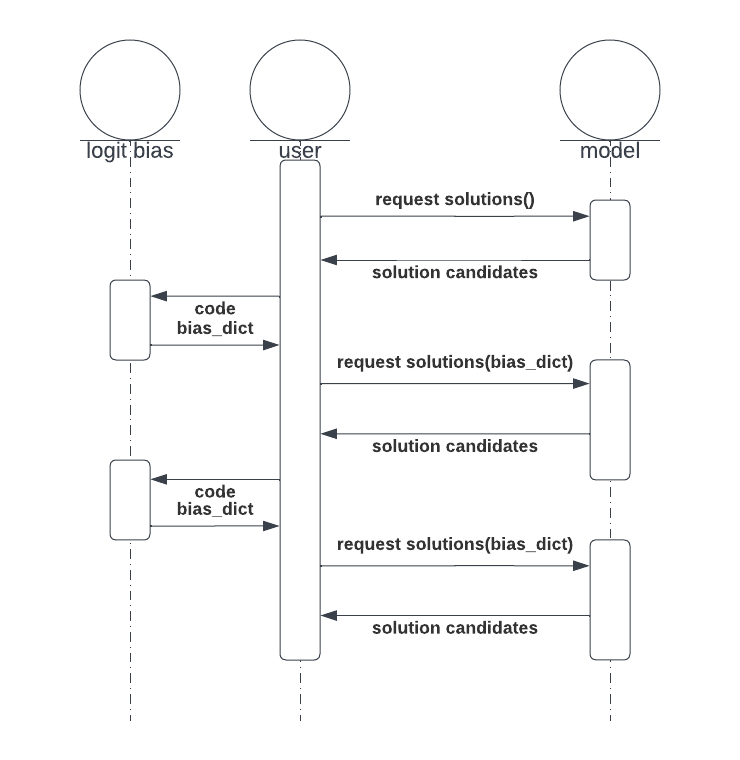}\vspace{-3mm}
    \caption{n\_k\_different prompt with Logit Bias; UML sequence diagram.}
    \label{fig:regen_logit_bias}
\end{figure}

The n\_k\_different prompt is similar to the n\_different prompt, but instead of asking the model to generate all $n$ solutions at once, $k$ solutions are asked to be generated ($n=10$ and $k=3$ are used). This is repeated until all $n$ solutions have been generated, where the last iteration asks for the remaining number of solutions. 
The advantage of this technique over n\_different is the ability to use the logit bias parameter as seen in Chung et al.~\cite{chung2023increasing} between API calls (see Figure~\ref{fig:regen_logit_bias}).
 

\paragraph{Prompt Classification}

As mentioned earlier, each problem description comes with some example test cases, making all prompts few-shot/n-shot prompts~\cite{NEURIPS2020_1457c0d6} with regard to the goal of achieving correctness. However, in terms of the task of increasing diversity, we consider all prompts to be classified as zero-shot, as no examples relating to diversity are given in the prompts.


\section{Experimental Study}

In the following, we report our experimental results. In the first set of experiments, where we request diverse code solutions from foundation models, we change one aspect at a time. In the second set of experiments, we combine those aspects that represented the best diversity-correctness trade-offs in an attempt to create even further trade-offs.




\paragraph{Implementation Details} 
For security reasons, since we execute code generated by a foundation model, we follow the approach of the HumanEval study~\cite{chen2021codex} and execute tests in a sand-boxed Ubuntu virtual machine. 

Because models may output more than one function as part of the solution, this raises the problem of determining which function is the ``solution'' function. We solve this by checking each function until the test cases are passed or all functions are exhausted (in which case the candidate is flagged as having failed). 

For practical reasons, we give each model only one attempt at producing a correct set of solutions for a task, even though foundation models can be non-deterministic in their behaviour. To mitigate this, we aggregate results over all 164 tasks and test each diversification approach in isolation.



\subsection{Results: One Technique at a Time}\label{sec:oneatatime}

In this part of the investigation, we consider gpt-3.5 as the starting point of our investigations and then vary one aspect at a time. In total, this results in 20 different configurations (see Table~\ref{tab:approaches}).

\renewcommand{\cellalign}{cl}
\setlength{\tabcolsep}{2pt}\def\arraystretch{1.0}
\begin{table}
\caption{Approaches investigated to request diverse code solutions. A0 represents the starting point of our investigations, as it uses gpt-3.5 with default parameters.}
\label{tab:approaches}\vspace{-3mm}
\begin{tabular}{lllccccc}
\toprule
              & model         & prompt        & temperature & top\_p & \makecell{frequency\\penalty} & \makecell{presence\\penalty} & \makecell{logit\\bias}         \\
    \midrule
A0             & gpt-3.5 & n\_different   & 1           & 1      & 0                 & 0                & -               \\\midrule
A1               & gpt-4         & n\_different   & 1           & 1      & 0                 & 0                & -               \\\midrule
A2              & gpt-3.5 & regen    & 1           & 1      & 0                 & 0                & -               \\
A3               & gpt-3.5 & n\_k\_different\hspace{-3mm} & 1           & 1      & 0                 & 0                & -               \\\midrule
A4         & gpt-3.5 & n\_different   & 0.3         & 1      & 0                 & 0                & -               \\
A5          & gpt-3.5 & n\_different   & 0.7         & 1      & 0                 & 0                & -               \\
A6          & gpt-3.5 & n\_different   & 0.9         & 1      & 0                 & 0                & -               \\
A7          & gpt-3.5 & n\_different   & 1.3         & 1      & 0                 & 0                & -               \\\midrule
A8            & gpt-3.5 & n\_different   & 1           & 0.2    & 0                 & 0                & -               \\
A9               & gpt-3.5 & n\_different   & 1           & 0.4    & 0                 & 0                & -               \\
A10              & gpt-3.5 & n\_different   & 1           & 0.6    & 0                 & 0                & -               \\
A11              & gpt-3.5 & n\_different   & 1           & 0.8    & 0                 & 0                & -               \\\midrule
A12 & gpt-3.5 & n\_different   & 1           & 1      & -2.0                & 0                & -               \\
A13   & gpt-3.5 & n\_different   & 1           & 1      & -0.5              & 0                & -               \\
A14   & gpt-3.5 & n\_different   & 1           & 1      & 0.5               & 0                & -               \\
A15   & gpt-3.5 & n\_different   & 1           & 1      & 2.0                 & 0                & -               \\\midrule
A16  & gpt-3.5 & n\_different   & 1           & 1      & 0                 & -2.0               & -               \\
A17    & gpt-3.5 & n\_different   & 1           & 1      & 0                 & -0.5             & -               \\
A18    & gpt-3.5 & n\_different   & 1           & 1      & 0                 & 0.5              & -               \\
A19    & gpt-3.5 & n\_different   & 1           & 1      & 0                 & 2.0                & -               \\\midrule
A20        & gpt-3.5 & n\_k\_different\hspace{-3mm} & 1           & 1      & 0                 & 0                & \hspace{-3mm}\makecell{Chung\\et~al.~\cite{chung2023increasing}}\\
\bottomrule\vspace{-2mm}
\end{tabular}
\end{table}

\begin{figure*}
    \centering\vspace{-3mm}%
    \caption{Code similarity and correctness for all 20 approaches from Section~\ref{sec:oneatatime} and for the 2 approaches from Section~\ref{sec:combination}. Mean similarity scores are reported across all 164 HumanEval tasks.
    The red star shows the starting point of our investigations (A0).\\
    Sub-figures (a) and (b): 
    the purple/green triangles represent the Pareto fronts (from left to right): purple A15, A14, A1, A20, green A15, A1, A20. 
    The light-green diamonds represent A21 and A22 in Section~\ref{sec:combination}. \\
    Sub-figure (c): we show the correlation of the two clone detection approaches (Spearman correlation coefficient 0.993).}\vspace{2mm}
    \centering\hspace{8mm}(a)\hspace{52mm}(b)\hspace{53mm}(c)\\
    \includegraphics[width=5.5cm,height=4.4cm,trim={10 20 5 22},clip]{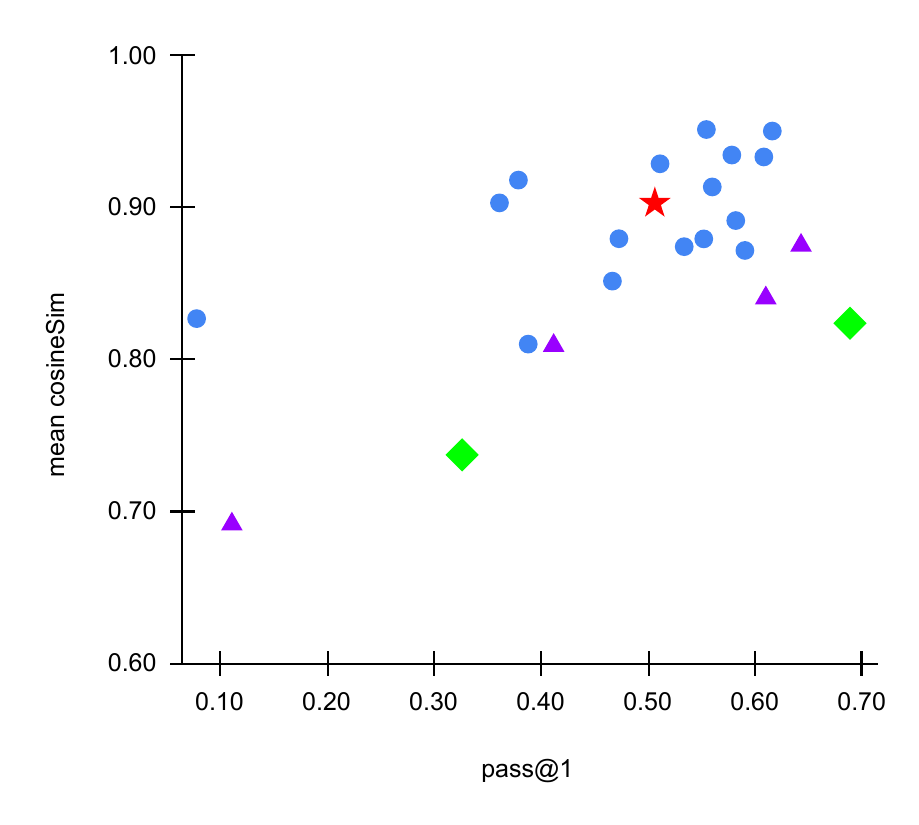}
    \includegraphics[width=5.5cm,height=4.4cm,trim={10 20 5 22},clip]{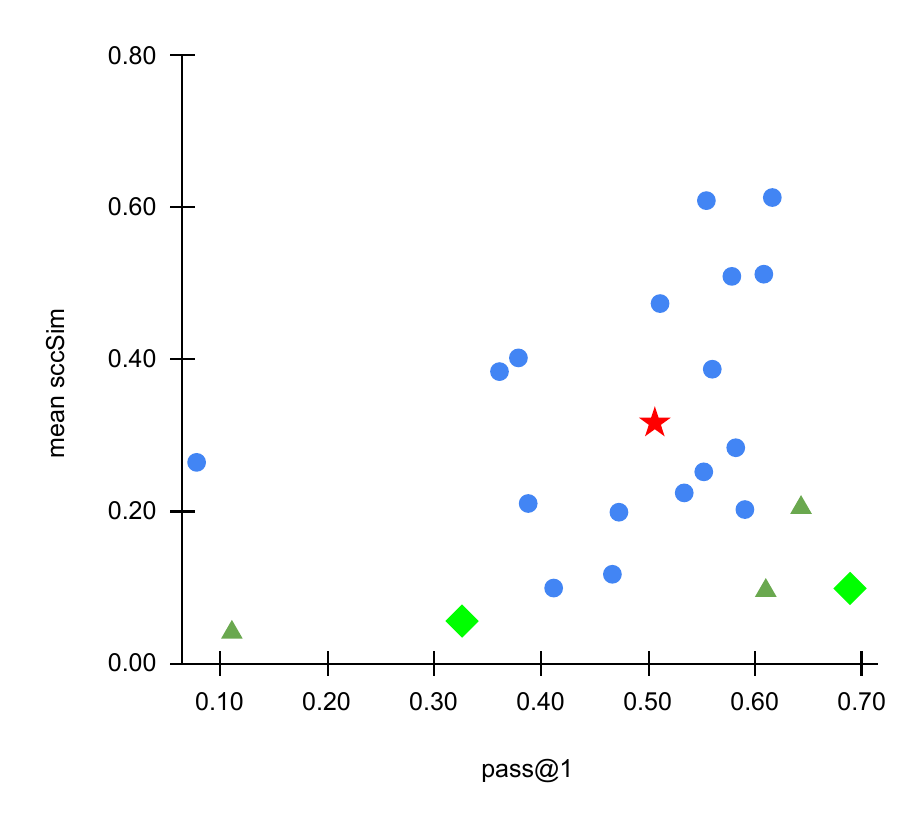}
    \includegraphics[width=5.5cm,height=4.4cm,trim={10 20 5 22},clip]{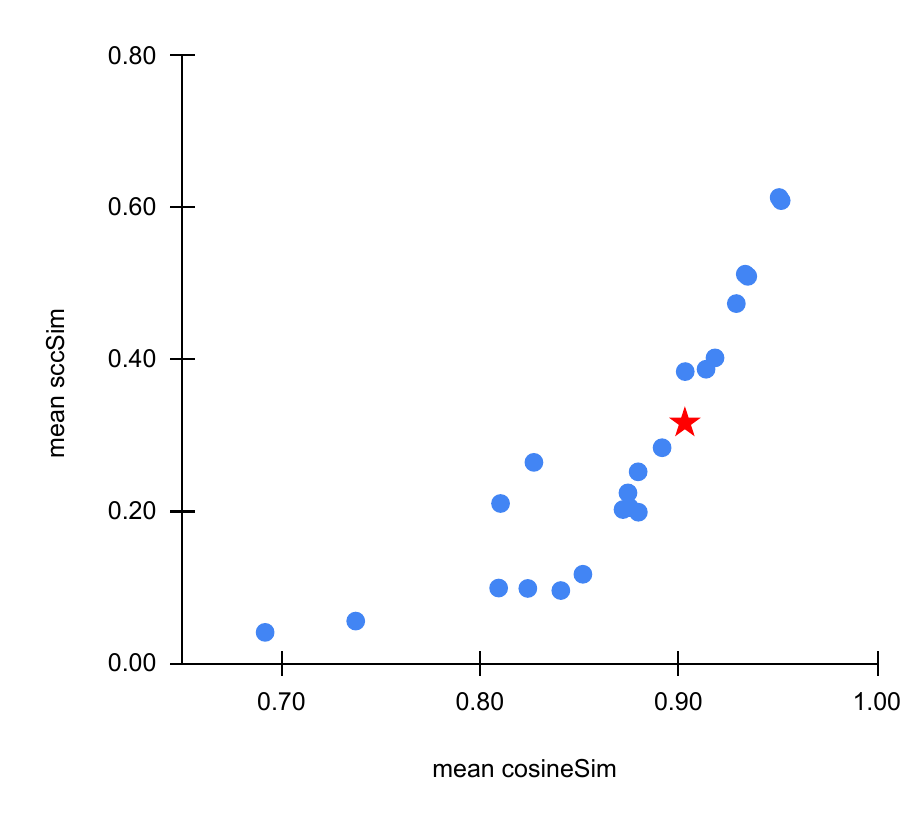}
    \label{fig:results}\vspace{-2mm}
\end{figure*}

We show the results in Figure~\ref{fig:results}; the data is also available online~\cite{spreadsheet} (anonymized). We highlight the Pareto front in either violet or green, i.e., the solutions that represent the best-possible trade-offs between solution correctness (as expressed by the pass@1 scores, where larger values are better) and the diversity metrics cosineSim and sccSim, where smaller values are better.

Interestingly, in both cases, A15 and A20 form the extreme solutions on the Pareto front: A15 is the solution with the best diversity among its outputs (0.69 mean cosineSim, 0.04 mean sccSim) while sporting a very low pass@1 rate (0.11); A20 has the highest pass@1 rate (0.69) while sporting a relatively low diversity (0.88 mean cosineSim, 0.20 mean sccSim). In contrast to these two solutions, the default configuration A0 (red star) does not represent a trade-off that should be considered, as it is worse than both approaches in terms of diversity.

Remarkably, our pass@1 scores are much higher than those observed in the HumanEval study~\cite{chen2021codex}: theirs range (across a range of approaches) from 0.01 to 0.28 (while focussing only on correctness), and our scores range from 0.08 to 0.69 (while considering diversity at the same time). However, we cannot claim that this is due to our diversity focus, as our scores are aligned with those listed by Achiam et al.~\cite{openai2023gpt4}, who report pass@1 scores of 0.48 for gpt-3.5 and 0.67 for gpt-4. For custom approaches to HumanEval, \cite{humanevallist} lists approaches that achieve pass@1 scores of up to 0.94.

Regarding the operational parameters, 
we make the following observations. 
First, the diversity increases (and pass@1 decreases) consistently as the temperature increases. 
Second, for presence\_penalty, we observe that the diversity increases with increasing penalty value; however, the pass@1 score does not follow an easily identifiable trend. 
Third, for frequency\_penalty, the default settings appear to be those with the worst diversity and best pass@1 scores; towards either extreme (-2.0 and 2.0), diversity increases and pass@1 scores worsen. 
Lastly, for top\_p, we did not observe notable trends. 

On the topic of diversity assessments, we note that both metrics --- although structurally very different --- are highly correlated in our study, with a Spearman correlation coefficient of 0.993 (see Figure~\ref{fig:results}(c)).


\subsection{Results: Combination of Techniques}\label{sec:combination}

Next, we combine aspects of the solutions that represent the Pareto front with the goal to create approaches that represent further Pareto-optimal trade-offs between diversity and correctness. 

The approaches A21 and A22 are listed in Table~\ref{tab:furtherapproaches}: both employ
frequency penalty (values 0.5 (A21) and 2.0 (A22), due to A14/A15), 
gpt-4 (due to A1) and
logit bias according to Chung et~al.~\cite{chung2023increasing} (due to A20).

The results are also shown in Figure~\ref{fig:results}, as light-green diamonds. We observe that both recombinations result in new, Pareto-optimal trade-offs. A22 even achieves the highest pass@1 score across all our approaches (0.69).

\begin{table}\def\arraystretch{0.95}
\caption{Approaches investigated based on combining aspects of Pareto optimal trade-offs (see Figure~\ref{fig:results}).}
\label{tab:furtherapproaches}\vspace{-3mm}
\begin{tabular}{llcccccc}
\toprule
              & model         & prompt        & temperature & top\_p & \makecell{frequency\\penalty} & \makecell{presence\\penalty} & \makecell{logit\\bias}         \\
    \midrule
A21    & gpt-4 & n\_k\_different\hspace{-3mm}  & 1           & 1      & 0.5                 & 0                & \hspace{-3mm}\makecell{Chung\\et~al.~\cite{chung2023increasing}}\\
A22        & gpt-4 & n\_k\_different\hspace{-3mm} & 1           & 1      & 2.0                 & 0                & \hspace{-3mm}\makecell{Chung\\et~al.~\cite{chung2023increasing}}\\
\bottomrule
\end{tabular}\vspace{-2mm}
\end{table}


\vspace{-1mm}
\section{Summary and Future Work}

The ability of AI foundation models to produce a large number of responses for a single prompt is a significant asset, especially in software engineering, where diverse code solutions are essential for addressing different contexts and fostering creativity. In this paper, we have explored strategies to leverage foundation models to generate code solutions that are both diverse and correct. We achieved this by (1) altering the operational parameters of the models and (2) experimenting with various prompting strategies. Our research demonstrates that there is a trade-off between creativity and correctness, and we have shown that by recombining aspects of Pareto-optimal trade-offs, we can create new, effective trade-offs.

Moving forward, our focus will be on two main areas. First, our aim is to inform practitioners about effective prompting strategies that can yield diverse and correct solutions, such as iterative prompting and adjusting the frequency penalty. Second, we plan to automate this process. Applying the DUO principle~\cite{Agrawal2020duo}, we will use optimisation to fine-tune these strategies to achieve the best trade-offs. This approach will enable us to make the most of existing models for creating code solutions that are both creative and correct. Our initial efforts in manual recombination show the potential of this approach.

\bibliographystyle{ACM-Reference-Format}\bibliography{forge24}\end{document}